\documentstyle[twocolumn,prl,aps]{revtex}
\begin{document}
\def \beq{\begin{equation}}
\def \eeq{\end{equation}}
\def \beqarr{\begin{eqnarray}}
\def \eeqarr{\end{eqnarray}}

\draft

\twocolumn[\hsize\textwidth\columnwidth\hsize\csname @twocolumnfalse\endcsname

\title{
Nondissipative Drag Conductance as a Topological Quantum Number
}

\author{Kun Yang}

\address{
National High Magnetic Field Laboratory and Department of Physics,
Florida State University, Tallahassee, Florida 32310$^*$
}
\address{
and
Condensed Matter Physics 114-36, California Institute of Technology,
Pasadena, California 91125
}

\author{A. H. MacDonald}

\address{
Physics Department, Indiana University, Bloomington, Indiana 47405
}

\date{\today}
\maketitle

\begin{abstract}

We show in this paper that the boundary condition averaged 
nondissipative drag conductance of two coupled mesoscopic rings
with no tunneling,
evaluated in a particular many-particle eigenstate,
is a topological invariant characterized by a Chern integer. 
Physical implications of this observation are discussed.

\end{abstract}
\pacs{Pacs: 73.40.H, 73.61.G}
]

The interplay between electron-electron interactions and a random 
disorder potential gives
rise to fascinating new physics in low-dimensional electron systems. 
Ordinary transport measurements continue to be one of the most important 
experimental probe of low-dimensional electron systems, but are 
usually more sensitive to disorder than to 
interactions.  (The fractional quantum Hall effect is one exception.)
Recently, technological advances have made a new class of transport
experiments, ``drag measurements",  possible. 
These probe electron-electron interaction
directly, and thus bring the interplay between disorder and 
interactions into sharp focus\cite{gramila,rojo} and stimulated extensive
experimental\cite{expt} and theoretical\cite{theo} studies.

In a drag measurement\cite{gramila}, separate electrical contacts are made to 
electron gases in 2D layers or 1D wires or rings coupled only by 
electron-electron Coulomb or phonon-mediated interactions.
In a drag experiment  
one studies the charge response of one layer/wire to electric fields in a  
{\em different} layer/wire.
This response may be parametrized by a differential drag conductance $g_D$,
which is defined as the current change in one layer/wire due to 
a change in the voltage drop across the
second layer/wire:
\beq
g_D={dI_1\over dV_2}.
\eeq
Clearly, $g_D=0$ in the absence of interlayer/wire electron-electron 
interaction; thus
drag measurements probe the effects of electron-electron interactions directly.

In macroscopic systems at finite temperature ($T$), the drag response is
dominated by dissipative processes, and typically vanishes in the limit
$T\rightarrow 0$. Very recently the so called ``nondissipative drag", which
does not involve dissipative processes and can survive the $T\rightarrow 0$
limit, has attracted much attention\cite{rm,ulloa,canali,baker,rojo}.
This nondissipative drag effect is often discussed in the context of coupled
mesoscopic rings, in which the finite-size nature of the system
gives rise to phase coherence and to discrete energy levels.  
Nondissipative drag reflects the interaction induced dependence of
the persistent current\cite{reidel}
flowing in one ring on the magnetic flux threading the 
other ring. 

In most previous theoretical studies of drag (dissipative or nondissipative),
inter-layer/wire electron-electron interactions 
are treated as a weak perturbation\cite{zheng}. While such an approach
has yielded results that are mostly
in good qualitative agreements with experiments
thus far\cite{rojo}, it is also known that there exists cases in which 
strong inter-layer/wire electron-electron interactions can give rise to  
non-perturbative drag effects impossible to address using perturbation 
theory\cite{yang}. It would thus be very useful to develop theoretical 
methods in studies of drag that are non-perturbative in nature.
In this paper, we use the
topological approach to transport theory to establish some
exact results about nondissipative drag in coupled mesoscopic rings.
Specifically, we show that 
the boundary condition averaged nondissipative 
drag conductance $g_D$ of any non-degenerate
many-particle state is a measure of a topological property of this state,
and is quantized in units of $e^2/h$.

In the following, we first present a formal proof of the quantization
of $g_D$ in coupled rings. This analysis follows the topological path blazed by 
Thouless and co-workers\cite{tknn,niu} which has provided valuable
insight into the quantum Hall effect.
We then discuss the physical implications of our results.

Consider two coupled rings 
described by the following Hamiltonian:
\beq
H=\sum_{i,a}{\hbar^2\over 2m_e}\left(-i{\partial\over \partial x^a_i}
\right)^2+
\sum_{i,a}U_a(x^a_i)+{1\over2}\sum_{ijab}V_{ab}(x^a_i, x^b_j),
\eeq
where $m_e$ is the mass of the electron,
$a=1,2$ is the ring index, $x^a_i$ is the $x$ coordinate of the $i$th
electron in ring $a$, and $U$ and $V$ are some generic
ring-dependent one- and two-body 
potentials. The Kubo formula for the real part of the
{\em nondissipative}
drag conductance along the $\hat{x}$ direction
when the system is in eigenstate $\psi_m$ is\cite{niu,kubonote}
\beq
g_D^m={ie^2\hbar\over L_1L_2}\sum_{n\ne m}{\langle \psi_m|v_x^1|\psi_n\rangle
\langle \psi_n|v_x^2|\psi_m\rangle- c. c.\over
(E_n-E_m)^2},
\eeq
where $L_1$ and $L_2$ are the length of the two rings,
$c. c.$ stands for complex conjugation, and
\beq
v_x^a=\sum_i{\hbar\over m_e}(-i{\partial\over \partial x^a_i})
\eeq
is the velocity operator along $\hat{x}$ direction in layer $a$.
To proceed further, we impose periodic boundary conditions 
with twist angles $\varphi_1$ and $\varphi_2$, for 
rings 1 and 2 respectively\cite{fluxnote}:
\beqarr
\psi_n(\cdots, x_i^1+L_1,\cdots)=e^{i\varphi_1}\psi_n(\cdots, x_i^1,\cdots),
\nonumber\\
\psi_n(\cdots, x_i^2+L_2,\cdots)=e^{i\varphi_2}\psi_n(\cdots, x_i^2,\cdots).
\eeqarr
$g_D^m$ of course depends on $\varphi_1$ and $\varphi_2$. 
We now make 
the following unitary transformation:
\beq
\phi_n=\exp\left(-{i\over L_a}\sum_{i,a}\varphi_ax_i^a\right)\psi_n,
\eeq
so that $\phi_n$ is always periodic with no twist angles. 
Transforming the Hamiltonian accordingly using
\beq
-i{\partial\over \partial x^a_i}\rightarrow -i{\partial\over \partial x^a_i
}+{\varphi_a\over L_a},
\eeq
one obtains
\beq
\tilde{v}_x^a={L_a\over \hbar}{\partial \tilde{H}\over\partial \varphi_a},
\eeq
where $\tilde{H}$ is the transformed, $\varphi_a$ dependent Hamiltonian,
and $\tilde{v}$ is the transformed velocity operator. Thus
\beqarr
&g_D^m(\varphi_1,\varphi_2)
&={ie^2\over\hbar}\sum_{n\ne m}{\langle \phi_m|{\partial\tilde{H}\over\partial 
\varphi_1}|\phi_n\rangle
\langle \phi_n|{\partial\tilde{H}\over\partial 
\varphi_2}|\phi_m\rangle- c. c.
\over
(E_n-E_m)^2}\nonumber\\
&=&{ie^2\over\hbar}\left[\left\langle{\partial\phi_m\over \partial 
\varphi_1}\left|{\partial\phi_m\over \partial 
\varphi_2}\right\rangle-\left\langle{\partial\phi_m\over \partial 
\varphi_2}\right|{\partial\phi_m\over \partial 
\varphi_1}\right\rangle\right].
\eeqarr
Thus the boundary condition averaged nondissipative drag conductance is
\beq
\overline{g_D^m}={ie^2\over 2\pi h}\int\int_0^{2\pi}{d\varphi_1}
d\varphi_2
\left[\left\langle{\partial\phi_m\over \partial
\varphi_1}\left|{\partial\phi_m\over \partial
\varphi_2}\right\rangle
-\left\langle{\partial\phi_m\over \partial
\varphi_2}\right|{\partial\phi_m\over \partial
\varphi_1}\right\rangle\right].
\eeq
This is of precisely the same form as the boundary condition averaged 
expression for the Hall conductance\cite{niu}. 
As in the quantum Hall case we can conclude that 
$\overline{g_D^m}=C(m)e^2/h$ 
is quantized in units of $e^2/h$, as long as state $m$ 
is not degenerate. The integer $C(m)$ is known as the Chern
integer or Chern number of state $m$. 

A few technical comments are in order.
i) While we have been discussing coupled mesoscopic rings with no transverse
cross sections, the above analysis can be applied to 2D systems or rings with
finite cross sections as well.
In the 2D case the conclusion is that the boundary condition averaged 
nondissipative drag
{\em conductance} of a given eigenstate is quantized, {\em not} the
drag {\em conductivity}. These two quantities have the same dimensionality 
in 2D,
but differ by a non-universal geometric factor. This is somewhat counter
intuitive, since one usually views the {\em conductivity} as an intrinsic 
property of the system, and the {\em conductance} as geometry dependent.
ii) In the preceding derivation we have simplified the discussion by 
using a kinetic energy term in the Hamiltonian that is
Galilean invariant and has no magnetic field.  
It is readily shown, however, that the result holds for 
any band structure and any magnetic field strength.
iii) In the two-dimensional case, the same analysis can be made and
the same conclusion reached for
the Hall drag conductance. 
iv) Our analysis breaks down when electrons are allowed to tunnel from 
one ring/layer to the other\cite{oreg}. 
This is because in the presence of tunneling,
the electrons do not have a well defined ring/layer index, and it is no longer 
possible to assign different boundary condition angles to different 
rings/layers.
v) In the derivation we need to assume the state $m$ under study is not 
degenerate with any other state for {\em any} $\varphi_1$ and $\varphi_2$.
This is the generic situation in the presence of randomness, 
since one needs to tune three parameters in a
Hermitian matrix to make two states degenerate.
However, in certain cases
there is a global degeneracy in the thermodynamic limit, which is of 
topological origin\cite{wen}.
In such cases the drag conductance is specified by the 
the average of the Chern numbers of the
degenerate ground states\cite{note2}, and can be a fractional multiple
of $e^2/h$.
This is the way in which fractional quantizations of ordinary
and Hall drag\cite{yang} conductances occur.

We now turn to a discussion of the physical implications of our results.
In the quantum Hall context, the identification of Hall conductance with a 
topological quantum number has lead to very important and fruitful insight
into the basic physics\cite{tknn,niu}. Here we attempt to analyze the physics
of nondissipative drag in light of this formally rather similar identification.
In the quantum Hall case, such an identification leads to the quantization of
the Hall conductance when the ground state of the system is separated from 
excited states by a real or mobility gap, as in the presence of such a gap,
(i) the system (and therefore the quantum number)
is very stable against perturbations, an essential ingredient 
for quantization; (ii) the boundary effect becomes very weak for sufficiently
large system size and thus the Hall conductance becomes {\em independent} of
boundary condition, leading to quantization of Hall conductance {\em without}
boundary condition averaging. Drag measurements, on the other hand, are 
usually performed in metallic systems with no gap in the thermodynamic limit;
for mesoscopic rings, the energy spacing between discrete levels are very
sensitive to system size and other perturbations. Thus we expect the drag
Chern number $C(m)$ to fluctuate from state to state, and $g_D^m$ to be 
sensitive to boundary conditions in spite of the quantization of
its boundary condition averaged value.  
Indeed, in the absence of an external field, 
it follows from time-reversal invariance that 
$g_D = 0 $ at $\phi_1 = \phi_2=0$ so that a non-zero
value of $\overline{g_D}$ {\em requires} strong sensitivity of 
the drag to flux phases.

We do, however, know one case in which the the drag conductance
{\em is} quantized.
This is the case where there is no interlayer interaction ($V_{12}=0$),
and $g_D^m$ or $C(m)$ is {\em identically zero}. 
Now let us turn on $V_{12}$ adiabatically, and focus on the ground state and 
its Chern number $C(0)$. Since $C(0)$ is quantized, 
it cannot change continuously with
$V_{12}$; it can only change when the ground state becomes degenerate with some
other state at a particular boundary condition and strength of $V_{12}$,
a situation which will be
called level crossing from now on\cite{note3}.
Thus $\overline{g_D}$ must remain
zero at $T=0$ even when the inter-ring interaction is turned on, as long as
it does not induce a level crossing of the ground state. Also from numerical
studies of Chern numbers in the quantum Hall\cite{huo} and other\cite{sheng}
contexts, we expect $g_D$ to be very close to $0$ with fluctuations small
compared to $e^2/h$, unless the system is very close to a level crossing. We
refer to this regime as the ``weak drag" regime. In this regime the magnitude
of $g_D$ is small compared to $e^2/h$, and the ground state of the system is
perturbatively connected the decoupled ring ground state, thus the 
inter-ring interaction can be treated perturbatively.

Stronger inter-ring interaction, on the other hand, can induce a ground state
level crossing, by which means the ground state can acquire a nonzero Chern
number $C(0)$. In this case we expect the magnitude of $g_D$ to be of order
$e^2/h$, and refer this to be the ``strong drag" regime. In this case the
ground state is {\em not} perturbatively 
connected the decoupled ring ground state,
and one can no longer treat inter-ring interaction as a perturbation.
We also expect strong fluctuations of $g_D$ in this regime, as a non-zero
Chern number is usually dominated by contributions in a small region in the
boundary condition space where two neighboring states come close together in
energy, in the boundary condition averaging process\cite{huo,sheng}. 

In the topological approach adopted here, one relates $g_D$ to the response of
the wave function of the system to the change of boundary conditions. For such
a response to be significant, the wave function must be delocalized. Disorder,
on the other hand, tends to localize the wave function and thus suppress such
response and therefore $g_D$. This observation is of course not surprising and
consistent with a recent study of nondissipative drag\cite{baker}. In the 
numerical studies of quantum Hall Chern numbers\cite{huo,sheng}, it was found
that for localized states, disorder suppresses Chern numbers much 
more strongly 
than other measures of localization (like longitudinal conductivity) in finite
size systems. This is related to the fact that Chern numbers are quantized, 
while other quantities can change continuously with disorder. It would be
interesting if the same phenomena occurs for drag Chern numbers; if so this
would suggest that disorder will suppress $g_D$ more strongly than the 
ordinary conductance in mesoscopic rings.

Quantized drag in coupled ring systems has an interesting 
physical interpretation.  Consider the {\em additional} current induced in
the first ring by a time-dependent magnetic flux threading a second ring.
The electric field in ring 2, $E_2$, is related to the 
time-derivative of the threading flux by $E_2 = \dot \phi_2/(c L_2)$ 
where $L_2$ is the perimeter of ring 2.  The additional
current induced in ring
1 is then $ \delta I_1 = C(m) (e^2/h) (\dot \phi_2 /c) $.  It follows
that the additional number of electrons flowing past a reference point in ring 
is $C(m)$ for each quantum of flux threading ring 2 which must be
an integer. This is, of course, very
similar in spirit
to Laughlin's original explanation of Hall conductance quantization 
based on gauge invariance\cite{laughlin}.

In summary, we have demonstrated that boundary condition averaged 
nondissipative drag 
conductance of individual eigenstates in a coupled bi-ring or bi-layer system 
is characterized by a topological quantum number. This 
quantization, and the resultant quantum numbers, provide a new way to view the
physics of nondissipative drag.  We expect that, as in the 
quantum Hall\cite{huo} and other contexts\cite{sheng}, 
numerical studies of these topological quantum 
numbers can shed considerable light on the physics of nondissipative drag. 

We would like to thank Jim Eisenstein, Yong Baek Kim, Ed
Rezayi, and Giovanni Vignale for useful discussions and correspondence. 
This work was supported by NSF grant No. DMR 99-71541, and the
Fairchild and Sloan Foundations (KY),
and by NSF grant No. DMR 97-14055 (AHM).
Part of the work was performed at ITP at Santa Barbara during
the program on {\em Disorder and Interactions in Quantum Hall and
Mesoscopic Systems}, which was supported by NSF PHY 94-07194.


\begin{references}
\bibitem{kubonote} 
In the complete expression of the Kubo formula for $g_D$, there
are also terms proportional to $\delta(E_m-E_n)$ describing transitions 
between different eigenstates induced by the external perturbation. Here we
neglect these terms on the ground that states are generically non-degenerate
in coherent mesoscopic systems. Those terms, however, are what gives rise
to dissipative drag response and must be retained in their studies. 
There is also an imaginary part of $g_D$
which is singular in the DC limit (proportional 
to to $1/i\omega$ where $\omega$ is the frequency), but it averages to zero
upon boundary condition condition averaging (see later), we thus neglect them 
here.
\bibitem{fluxnote} Physically such twisted boundary conditions can be induced
by inserting flux $\Phi_a={\varphi_a\over 2\pi}\Phi_0$ to ring $a$, where
$\Phi_0=hc/e$ is the flux quantum.
\bibitem{note2} The degeneracy becomes exact only in the thermodynamic
limit. In a finite but arbitrarily large system, there are small energy
differences among these states so that Chern numbers are well defined.
\end{references}
\end{document}